\documentclass[11pt]{article}
\usepackage{graphicx}
\usepackage{cite}
\usepackage{amssymb}
\usepackage{epstopdf}
\usepackage{hep}
\usepackage{feynmp}
\usepackage{multicol}
\usepackage{simplewick}
\usepackage{appendix}
\usepackage{subfigure}
\usepackage{color}
\usepackage{ulem}
\newcommand{\fig}[2]{\scalebox{#1}{\includegraphics{#2}}}
\newcommand{\eins}{\mbox{$1 \hspace{-1.0mm} {\bf l}$}}

\newcommand{\Sp}{S_\perp}
\newcommand{\vb}{\biggl|}

\allowdisplaybreaks

\begin{document}
\title{{\bf Longitudinal-transverse double-spin asymmetries in single-inclusive leptoproduction of hadrons}}

\author{K.~Kanazawa$^{1}$, A.~Metz$^{1}$, D.~Pitonyak$^{2}$, and M.~Schlegel$^{3}$
 \\[0.3cm]
{\normalsize\it $^1$Department of Physics, SERC,
  Temple University, Philadelphia, PA 19122, USA} \\[0.15cm]
{\normalsize\it $^2$RIKEN BNL Research Center,
                 Brookhaven National Laboratory,
                 Upton, New York 11973, USA} \\[0.15cm]
{\normalsize\it $^3$Institute for Theoretical Physics, T\"ubingen University,} \\ 
{\normalsize\it Auf der Morgenstelle 14, D-72076 T\"ubingen, Germany}            
}

\date{\today}
\maketitle

\begin{abstract}
\noindent
We analyze the longitudinal-transverse double-spin asymmetry in lepton-nucleon collisions where a single hadron is detected in the final state, i.e., $\vec{\ell}\,N^\uparrow \rightarrow h\,X$.  This is a subleading-twist observable in collinear factorization, and we look at twist-3 effects in both the transversely polarized nucleon and the unpolarized outgoing hadron.  Results are anticipated for this asymmetry from both HERMES and Jefferson Lab Hall A, and it could be measured as well at COMPASS and a future Electron-Ion Collider.  We also perform a numerical study of the distribution term, which, when compared to upcoming experimental results, could allow one to learn about the ``worm-gear''-type function $\tilde{g}(x)$ as well as assess the role of quark-gluon-quark correlations in the initial-state nucleon and twist-3 effects in the fragmenting unpolarized hadron.
\end{abstract}

%
%
\section{Introduction} \label{s:intro}
Hadrons, the strongly interacting particles that comprise almost all of the visible matter in the universe, have been shown to possess a complex inner-structure that goes beyond a simple quark picture.  For example, experimental results in the 1970s on transverse single-spin asymmetries (SSAs)~\cite{Bunce:1976yb} revealed the crucial role that quark-gluon-quark correlations could play in hadrons~\cite{Efremov:1981sh,Qiu:1991pp,Qiu:1998ia}.  This is a consequence of the fact that such observables are twist-3 effects. Much work over the last 40 years has been performed in the study of transverse SSAs from both the experimental (see, e.g.,~\cite{Bunce:1976yb,Adams:1991rw,Krueger:1998hz,Adams:2003fx,Adler:2005in,Lee:2007zzh,:2008mi,Airapetian:2009ab,Adamczyk:2012xd,Bland:2013pkt,Airapetian:2013bim,Allada:2013nsw,Katich:2013atq,Adare:2013ekj}) and theoretical (see, e.g.,~\cite{Efremov:1981sh,Qiu:1991pp,Qiu:1998ia,Kouvaris:2006zy,Kang:2011hk,Anselmino:1994tv,Anselmino:1998yz,Anselmino:2005sh,Anselmino:2012rq,Anselmino:2013rya,Kanazawa:2000hz,Eguchi:2006qz,Koike:2009ge,Kanazawa:2010au,Kanazawa:2011bg,Beppu:2013uda,Kang:2010zzb,Schlegel:2012ve,Metz:2012ui,Metz:2012ct,Kanazawa:2013uia,Kanazawa:2014dca,Kanazawa:2014nea}) sides.  In addition, one also has twist-3 double-spin asymmetries (DSAs), namely those where one particle is longitudinally polarized and the other is transversely polarized.  We will denote these by $A_{LT}$. The classic process for which this effect has been analyzed is $A_{LT}$ in inclusive deep-inelastic lepton-nucleon scattering (DIS) (see~\cite{Posik:2014usi} for recent experimental results on this observable).  In that case the entire result can be written in terms of the collinear twist-3 function $g_{T}(x)$.  Furthermore, this asymmetry has been studied in the Drell-Yan process involving two incoming polarized hadrons~\cite{Jaffe:1991kp,Tangerman:1994bb,Koike:2008du,Lu:2011th}; in inclusive lepton production from $W$-boson decay in proton-proton scattering~\cite{Metz:2010xs}; for jet production in lepton-nucleon collisions~\cite{Kang:2011jw}; and for direct photon production~\cite{Liang:2012rb}, jet/hadron production~\cite{Metz:2012fq}, and $D$-meson production~\cite{Hatta:2013wsa} in nucleon-nucleon collisions.

In this paper we analyze $A_{LT}$ in lepton-nucleon collisions where a single hadron is detected in the final state, i.e., $\vec{\ell}\,N^\uparrow \rightarrow h\,X$.  Transverse SSAs in single-inclusive leptoproduction processes have received some theoretical attention lately~\cite{Anselmino:2009pn,Kang:2011jw,Anselmino:2014eza,Gamberg:2014eia} due to recent experimental results from HERMES~\cite{Airapetian:2009ab} and Jefferson Lab (JLab) Hall A~\cite{Allada:2013nsw} on these observables as well as the potential for COMPASS and a future Electron-Ion Collider (EIC) to make such measurements.  Data on longitudinal-transverse DSAs in $\vec{\ell}\,N^\uparrow \rightarrow h\,X$ are also anticipated from HERMES~\cite{Schnell:priv2014} and JLab Hall A~\cite{Meziani:priv2013,Jiang:priv2014}, and both COMPASS and an EIC~\cite{Boer:2011fh,Accardi:2012qut} could run such an experiment, too.  This work is, therefore, a timely endeavor. 

For the reaction considered, one can have twist-3 contributions from both the distribution (incoming nucleon) and the fragmentation (outgoing hadron) sides.  We compute both of these as well as provide numerical results for the distribution term based on known non-perturbative inputs.  As a consequence, one has the opportunity, through a comparison of this phenomenology with experiment, to learn about the ``worm-gear''-type function $\tilde{g}(x)$ (defined in Sec.~\ref{s:theory}) as well as the role of quark-gluon-quark correlations in the nucleon and twist-3 effects in the fragmenting hadron.  There has been quite some interest in $\tilde{g}(x)$ over the years.  From the experimental side, one can access a transverse momentum dependent (TMD) analogue of $\tilde{g}(x)$ (denoted $g_{1T}(x,\vec{k}_\perp^2)$)\footnote{A precise relation between $\tilde{g}(x)$ and $g_{1T}(x,\vec{k}_\perp^2)$ will be stated in Sec.~\ref{s:theory}.} from the longitudinal-transverse $\cos(\phi_h-\phi_S)$ azimuthal asymmetry in semi-inclusive deep-inelastic scattering (SIDIS). JLab Hall A obtained the first results on this asymmetry~\cite{Huang:2011bc} and COMPASS also has preliminary data on this modulation~\cite{Parsamyan:2013fia}.  From the theoretical side, $\tilde{g}(x)$ has been looked at in a Wandzura-Wilczek (WW) -type approximation~\cite{Kotzinian:2006dw,Avakian:2007mv,Kang:2011jw}, in spectator models~\cite{Jakob:1997wg,Bacchetta:2008af}, in quark models~\cite{Pasquini:2008ax,Zhu:2011zza}, and in Lattice QCD~\cite{Hagler:2009mb}.  The analysis presented here can also contribute to our knowledge of $\tilde{g}(x)$.

The paper is organized as follows: in Sec.~\ref{s:theory} we set up the framework of the calculation and outline the derivation of our result; in Sec.~\ref{s:num} we discuss our numerical study and show plots for HERMES, JLab, COMPASS, and EIC kinematics; finally in Sec.~\ref{s:sum} we summarize our work.  Some details on the frame-independence of our result is left for Appendix A.

%
%
\section{Theoretical framework and result for the cross section} \label{s:theory}
In this section we define the relevant non-perturbative twist-3 correlators and present some details of our computation of the leading-order (LO) double-spin dependent cross section for the process
\begin{equation}
\vec{\ell}(l,\lambda_\ell) + N^\uparrow(P,S_\perp) \rightarrow h(P_h) + X\,, \label{e:LTreac}
\end{equation}
where we will work in the lepton-nucleon center-of-mass ($cm$) frame with the nucleon moving along the $+z$-axis and the transverse momentum of the outgoing hadron, $\vec{P}_{h\perp}$, along the $+x$-axis.  The Mandelstam variables for the process are defined as $S = (P+l)^{2}$, $T = (P-P_h)^{2}$, and $U = (l-P_h)^{2}$, which on the partonic level give $\hat{s} = xS$, $\hat{t} = xT/z$, and $\hat{u} = U/z$.

We first start with the twist-3 functions for a transversely polarized nucleon.  For a detailed discussion of collinear twist-3 distribution correlators, see, e.g.,~\cite{Zhou:2009jm}.  Note that tri-gluon matrix elements will not enter at LO in this reaction, and so we will not discuss them here.  In the lightcone gauge with $A^{+} = 0$ we have the so-called ``F-type''  and ``D-type'' correlators~\cite{Qiu:1991pp},
\begin{align}
\!\int\!& \frac{d\xi^-}{2 \pi} \int\!\frac{d\zeta^-}{2\pi} e^{ix_{1}P^+ \xi^-} \!e^{i(x-x_1)P^+\zeta^-}\langle P,S_\perp | \bar{\psi}^q_j(0)  g F_\perp^{+ \mu}(\zeta^-)\psi^q_i(\xi^{-}) | P,S_\perp \rangle\nonumber\\
&= \frac{M}{2} \Big[ F^q_{FT}(x,x_1) \, \epsilon^{\mu\nu}_\perp S_{\perp \nu}\, \gamma^-
- G^q_{FT}(x,x_1) \, i S_\perp^\mu\, \gamma_5 \gamma^- \Big]_{ij}, \label{e:F-type}\\[0.35cm]
\!\int\! &\frac{d\xi^-}{2 \pi} \int\!\frac{d\zeta^-}{2\pi} \, e^{ix_{1}P^+ \xi^-} \, e^{i(x-x_1)P^+\zeta^-}\langle P,S_\perp | \bar{\psi}^q_j(0)  i D_\perp^\mu(\zeta^-)\psi^q_i(\xi^{-}) | P,S_\perp \rangle\nonumber\\
&= \frac{M}{2P^+} \Big[ F^q_{DT}(x,x_{1}) \, i\epsilon^{\mu\nu}_\perp S_{\perp\nu}\, \gamma^-
+G^q_{DT}(x,x_{1}) \, S_\perp^\mu\, \gamma_5 \gamma^- \Big]_{ij}\,, \label{e:D-type}
\end{align}
where $M$ is the nucleon mass, and $\epsilon_\perp^{\mu\nu}\equiv\epsilon^{-+\mu\nu}$ with $\epsilon^{0123} = +1$. We refer the reader to Ref.~\cite{Metz:2012fq} for the symmetry properties of and relations between these F-type and D-type functions.  One also needs the function $\tilde{g}(x)$,
\begin{eqnarray}
&& \int\! \frac{d\xi^-}{2 \pi} \, e^{ixP^+ \xi^-}
\langle P,S_\perp | \bar{\psi}^q_j(0)  
\hspace{-0.025cm}\bigg( iD_\perp^\mu(\xi^{-}) + g\int_{\xi^{-}}^\infty \!d\zeta^- F_\perp^{+\mu}(\zeta^-) \bigg)\hspace{-0.025cm}
\psi^q_i(\xi^{-}) | P,S_\perp \rangle
\nonumber\\
&& \hspace{0.5cm}
= \frac{M}{2} \Big[ \tilde{g}^q(x) \, S_\perp^\mu\, \gamma_5 \gamma^- \Big]_{ij}\,.
\end{eqnarray}
We mention that $\tilde{g}(x)$ is equivalent to the first $k_{\perp}$-moment of the TMD function $g_{1T}(x,\vec{k}_{\perp}^{2})$ (defined in~\cite{Boer:1997nt,Goeke:2005hb})~\cite{Zhou:2009jm},
\begin{equation}
\tilde{g}^q(x) =  g^{q(1)}_{1T}(x)\equiv \int\! d^{2}\vec{k}_\perp \frac{\vec{k}_\perp^2} {2M^2}\, g_{1T}^q(x,\vec{k}_\perp^2)\,.
 \label{e:g1T}
\end{equation}
The last function required is $g_{T}(x)$, given by~\cite{Mulders:1995dh}
\begin{equation}
\int\!\frac{d \xi^-} {2\pi}e^{ixP^+\xi^-}\langle P,S_\perp|\bar{\psi}^q_j(0)\psi^q_i(\xi^-)|P,S_\perp\rangle = \frac{M} {2P^{+}}\left[g^q_{T}(x)\,S_{\perp}^{\mu}\,\gamma_5\gamma_\mu\right]_{ij}\,.  \label{e:gT}
\end{equation}
It turns out that $g_{T}(x)$ can be related to the D-type functions through the QCD equation of motion (EOM)~\cite{Efremov:1981sh,Jaffe:1991kp}, 
\begin{equation} 
x\,g^q_{T}(x)=\int\! dx_{1}\, \left[G^q_{DT}(x,\,x_{1})-F^q_{DT}(x,\,x_{1})\right]\,, \label{e:EOMgT} 
\end{equation}
and to $G_{DT}(x,x_1)$ and the helicity distribution $g_1(x)$ (defined in, e.g., Ref.~\cite{Jaffe:1991kp}) through a Lorentz invariance relation (LIR)~\cite{Buchvostov:1984xf,Belitsky:1997ay,Kundu:2001pk,Eguchi:2006qz,Accardi:2009au},
\begin{equation}
g_T^q(x) = g^q_1(x) - \frac{2} {x}\int \!dx_1\,\frac{1} {\xi}\,G^q_{DT}(x,x_1)\,, \label{e:LIR}
\end{equation}
where $\xi = (x-x_1)/x$, and we understand $1/\xi$ to mean $PV(1/\xi)$.

We next look at the twist-3 fragmentation functions (FFs) for an outgoing unpolarized hadron. (Note that tri-gluon FFs only enter with transversely polarized hadrons.)  In the lightcone gauge with $A^- = 0$ one also has F-type and D-type functions,
\begin{align}
\sum_{X}\hspace{-0.55cm} \int\, \frac{1} {z}\!\int\! \frac{d\xi^{+}} {2\pi}\!\int\! \frac{d\zeta^{+}} {2\pi} &e^{i\frac{P_{h}^{-}} {z_{1}}\xi^{+}} e^{i\left(\frac{1} {z}-\frac{1} {z_1}\right)P_{h}^{-}\zeta^{+}} \langle 0|gF_{\perp}^{-\mu}(\zeta^{+})\psi^q_{i}(\xi^{+})|P_{h};X\rangle\langle P_{h};X|\bar{\psi}^q_{j}(0)|0\rangle \nonumber\\
&= -M_{h}\left[i\epsilon_{\perp}^{\mu\nu}\,\sigma_{\nu}^{\;\,+}\gamma_{5}\,\hat{H}^{h/q}_{FU}(z,z_{1})\right]_{ij},\label{e:F-typeFF} \\[0.2cm]
\sum_{X}\hspace{-0.55cm} \int \; \frac{1} {z}\int\! \frac{d\xi^{+}} {2\pi}\int\! \frac{d\zeta^{+}} {2\pi} &e^{i\frac{P_{h}^{-}} {z_{1}}\xi^{+}} e^{i\left(\frac{1} {z}-\frac{1} {z_1}\right)P_{h}^{-}\zeta^{+}}\langle 0|iD_{\perp}^{\mu}(\zeta^{+})\psi^q_{i}(\xi^{+})|P_{h};X\rangle\langle P_{h};X|\bar{\psi}^q_{j}(0)|0\rangle \nonumber\\
 &=  \frac{M_{h}} {P_{h}^{-}}\left[\epsilon_{\perp}^{\mu\nu}\,\sigma_{\nu}^{\;\,+}\gamma_{5}\,\hat{H}^{h/q}_{DU}(z,z_{1})\right]_{ij}, \label{e:D-typeFF}
\end{align}
where $M_h$ is the hadron mass.  We remark that these functions contain both real and imaginary parts, which we respectively indicate by $\hat{H}^{\Re}_{FU(DU)}(z,z_{1})$ and $\hat{H}^{\Im}_{FU(DU)}(z,z_{1})$.  We refer the reader to \cite{Metz:2012ct,Kanazawa:2014dca} for relations between the F-type and D-type functions. One also needs the function $\hat{H}(z)$, 
\begin{align}
\sum_{X}\hspace{-0.55cm}\int \; z\!&\int\!\frac{d\xi^{+}} {2\pi} e^{i\frac{P_{h}^{-}} {z}\xi^{+}}\langle 0 |\!\left( iD_\perp^\mu(\xi^{+}) + g\int_{\xi^{+}}^\infty\! d\zeta^+ F_\perp^{-\mu}(\zeta^+) \right)\hspace{-0.025cm}\psi^q_{i}(\xi^{+})|P_{h}; X\rangle \langle P_{h}; X|\bar{\psi}^q_{j}(0)|0\rangle \nonumber\\
&= -iM_{h}\left[\epsilon_{\perp}^{\mu\nu}\,\sigma_{\nu}^{\;\,+}\gamma_{5}\,\hat{H}^{h/q}(z)\right]_{ij}. \label{e:Hhatdef}
\end{align}
We mention that $\hat{H}(z)$ is equivalent to the first $p_{\perp}$-moment of the TMD Collins function $H_{1}^{\perp}(z,z^2\vec{p}_{\perp}^{\,2})$ (first considered in~\cite{Collins:1992kk})~\cite{Kang:2010zzb,Metz:2012ct,Yuan:2009dw},
\begin{equation}
\hat{H}^{h/q}(z) =  H_{1}^{\perp\,h/q(1)}(z)\equiv z^2\!\int\! d^{2}\vec{p}_\perp \frac{\vec{p}_\perp^{\,2}} {2M_h^2}\,H_1^{\perp\,h/q}(z,z^2\vec{p}_\perp^{\,2}) \,. \vspace{-0.1cm}
 \label{e:H1perp}
\end{equation}
The last two functions required are $H(z)$ and $E(z)$, given by~\cite{Mulders:1995dh}
\begin{eqnarray}
\sum_{X}\hspace{-0.55cm}\int \; z\! \int\frac{d\xi^{+}} {2\pi} e^{i\frac{P_{h}^{-}} {z}\xi^{+}}\!\langle 0 |\psi^q_{i}(\xi^{+})|P_{h}; X\rangle\langle P_{h}; X|\bar{\psi}^q_{j}(0)|0\rangle = \frac{M_{h}} {2P_{h}^{-}}\!\left[-i\epsilon_{\perp}^{\mu\nu}\,\sigma_{\mu\nu}\gamma_{5}\,H^{h/q}(z) + 2E^{h/q}(z)\cdot \eins\right]_{ij}\!.\nonumber\\
\end{eqnarray}
It turns out $H(z)$ and $E(z)$ can be related to the imaginary part and real part, respectively, of the D-type function through the QCD EOM,
\begin{align} 
H^{h/q}(z) = 2z^3\int\! \frac{d z_1} {z_1^2} \hat{H}^{h/q,\Im}_{DU}(z,z_{1})\,, \label{e:EOMH}\\[0.3cm]
E^{h/q}(z) = -2z^3\int\! \frac{d z_1} {z_1^2} \hat{H}^{h/q,\Re}_{DU}(z,z_{1})\,. \label{e:EOME}
\end{align}

With the relevant twist-3 functions for this reaction now in hand, we proceed to the derivation of the double-spin dependent differential cross section $d\sigma_{LT}(\lambda_\ell,\vec{S}_\perp)$.  This will be used to calculate $A_{LT}$, defined as 
\begin{equation}
A_{LT} \equiv \frac{\Big\{\!\left[d\sigma_{LT}(+,\uparrow_x) - d\sigma_{LT}(-,\uparrow_x)\right]-\left[d\sigma_{LT}(+,\downarrow_x) - d\sigma_{LT}(-,\downarrow_x)\right]\!\Big\}} {4\,d\sigma_{unp}}\,, \label{e:ALT}
\end{equation}
where $+$ ($-$) indicates a lepton with positive (negative) helicity, $\uparrow_x$ ($\downarrow_x$) designates a nucleon with transverse spin along the $+x$ ($-x$) -axis, and $d\sigma_{unp}$ is the unpolarized cross section.  For the distribution term, the calculation follows along the lines of Refs.~\cite{Kang:2011jw,Liang:2012rb,Metz:2012fq}.  Specifically, one can modify the computation of the $qq^\prime\to qq^\prime$ channel in $\vec{p}\,p^\uparrow\to h\,X$~\cite{Metz:2012fq} to take into account the fact that photons do not carry color.  The result reads 
\begin{align}
&\hspace{-0.3cm}\frac{P_h^0\,d\sigma_{LT}^{Dist}(\lambda_\ell,\vec{S}_\perp)} {d^3\vec{P}_h} = -\frac{8\alpha_{em}^2} {S}\,M\,\vec{P}_{h\perp}\cdot\vec{S}_\perp\,\lambda_\ell\sum_q e_q^2\int_{z_{min}}^1\!\frac{dz} {z^3}\,\frac{1} {S+T/z}\,\frac{1} {x\hat{u}}\,D_1^{h/q}(z)\nonumber\\[0.05cm]
&\hspace{2cm}\times\,\Bigg\{\!\!\left(\tilde{g}^q(x)-x\frac{d\tilde{g}^q(x)} {dx}\right)\!\left[\frac{\hat{s}(\hat{s}-\hat{u})} {2\hat{t}^{\hspace{0.025cm}2}}\right]+x\,g_T^q(x)\left[\frac{\hat{u}} {2\hat{t}}\right]+ \int \!dx_1\,G_{DT}^q(x,x_1)\left[\frac{\hat{u}(\hat{s}-\hat{u})} {\xi\hat{t}^{\hspace{0.025cm}2}}\right]\!\!\Bigg\}\,,  \label{e:lNDist}
\end{align}
where $\alpha_{em}$ is the fine structure constant, $e_q$ is the (anti)quark charge with $\sum_q$ indicating a sum over both quarks and antiquarks, $x=-(U/z)/(S+T/z)$, and $z_{min}=-(T+U)/S$. We mention that one can obtain the cross section for $\vec{\ell}\,N^\uparrow\to jet\,X$~\cite{Kang:2011jw} from our result by making the replacement $D_1(z) \to \delta(1-z)$.\footnote{Note that the result in Ref.~\cite{Kang:2011jw} needs to be corrected:~the sign of the $g_T(x)$ term must be reversed, and the contribution containing $G_{DT}(x,x_1)$ must be added.}  Moreover, we performed our calculation in both lightcone gauge and Feynman gauge as well as in two different frames (lepton-nucleon $cm$ and nucleon-hadron $cm$) and found full agreement with Eq.~(\ref{e:lNDist}) in all cases.  We note that in order to show the frame-independence of the result, it was crucial to use the LIR (\ref{e:LIR}) as well as recognize that in the nucleon-hadron $cm$ frame, a contribution from $g_1(x)$ survives the asymmetry (since $A_{LT}$ in (\ref{e:ALT}) is defined with the transverse spin of the nucleon $\vec{S}_\perp$ in the lepton-nucleon $cm$ frame).  We discuss this equivalence between results in two different frames more in Appendix A.  For now we write down that version (i.e., using Eq.~(\ref{e:LIR}) in (\ref{e:lNDist})) of the formula here, as it will be useful in our numerical study presented in Sec.~\ref{s:num}:
\begin{align}
&\frac{P_h^0\,d\sigma^{Dist}_{LT}(\lambda_\ell,\vec{S}_\perp)} {d^3\vec{P}_h} = -\frac{8\alpha_{em}^2} {S}\,M\,\vec{P}_{h\perp}\cdot\vec{S}_\perp\,\lambda_\ell\sum_q e_q^2\int_{z_{min}}^1\!\frac{dz} {z^3}\,\frac{1} {S+T/z}\,\frac{1} {x\hat{u}}\,D_1^{h/q}(z)\nonumber\\[0.05cm]
&\hspace{2cm}\times\,\Bigg\{\!\!\left(\tilde{g}^q(x)-x\frac{d\tilde{g}^q(x)} {dx}\right)\!\left[\frac{\hat{s}(\hat{s}-\hat{u})} {2\hat{t}^{\hspace{0.025cm}2}}\right]+x\,g_T^q(x)\left[\frac{-\hat{s}\hat{u}} {\hat{t}^2}\right]+x\,g_1^q(x)\!\left[\frac{\hat{u}(\hat{s}-\hat{u})} {2\hat{t}^2}\right]\!\!\Bigg\}. \label{e:lNhXLT_new}
\end{align}

We now move on to the computation of the fragmentation piece.  This term has been calculated in the collinear twist-3 approach for the transverse SSA in $p^\uparrow\,p\to h\,X$~\cite{Kang:2010zzb,Metz:2012ct}, SIDIS~\cite{Yuan:2009dw,Kanazawa:2013uia}, and $\ell \,p^\uparrow\to h\,X$~\cite{Gamberg:2014eia}, where one has the transversity distribution $h_1(x)$ (defined in, e.g., Ref.~\cite{Jaffe:1991kp}) coupled to $\hat{H}(z)$, $H(z)$, and $\hat{H}_{FU}^{\Im}(z,z_1)$.  For the longitudinal-transverse case, one still has $h_1(x)$ entering on the side of the transversely polarized nucleon.  However, since we are considering a longitudinally polarized lepton, the relevant part of the leptonic tensor is purely imaginary, which brings an extra factor of $i$ into the derivation.  Consequently, the functions $\hat{H}(z)$, $H(z)$, and $\hat{H}_{FU}^{\Im}(z,z_1)$ do not contribute since the hadronic factors from these functions are purely real.  Therefore, one only has terms involving $E(z)$ and $\hat{H}_{FU}^{\Re}(z,z_1)$.\footnote{One could pick out the poles of $\hat{H}^{\Im}_{FU}(z,z_1)$ to produce another factor of $i$ from the hadronic side.  However, the partonic pole matrix elements for such functions have been shown to vanish~\cite{Gamberg:2008yt,Meissner:2008yf,Gamberg:2010uw}, which also follows from the universality of the Collins function~\cite{Metz:2002iz,Collins:2004nx,Yuan:2009dw}.  (For the same reason, one does not have the poles of $\hat{H}_{FU}^{\Re}(z,z_1)$ entering the fragmentation piece of transverse SSAs.)}  We find that through the QCD EOM relation (\ref{e:EOME}) the entire result for the fragmentation term can be written in terms of $E(z)$,
\begin{align}
&\hspace{-0.3cm}\frac{P_h^0\,d\sigma^{Frag}_{LT}(\lambda_\ell,\vec{S}_\perp)} {d^3\vec{P}_h} = -\frac{8\alpha_{em}^2} {S}\,M_h\,\vec{P}_{h\perp}\cdot\vec{S}_\perp\,\lambda_\ell\sum_q e_q^2\int_{z_{min}}^1\!\frac{dz} {z^3}\,\frac{1} {S+T/z}\,\frac{1} {zx\hat{t}}\,h_1^q(x)\,E^{h/q}(z)\left[-\frac{\hat{s}} {\hat{t}}\right].  \label{e:lNFrag}
\end{align}
As with the distribution term, we computed the fragmentation term in both lightcone gauge and Feynman gauge and in two different frames --- in all cases we obtained Eq.~(\ref{e:lNFrag}).  Note that the frame-independence of the fragmentation term is manifest without the need for a LIR.  

%
%
\section{Numerical results for $A_{LT}$} \label{s:num}
Here we present some numerical results for $A_{LT}$ in $\vec{e}\,N^\uparrow \to \pi\,X$, where $N = p,n$.  We emphasize that this is an exploratory study that will need guidance from experiment in order to learn anything quantitative.  We will only look at the distribution piece and use the form for this term given in Eq.~({\ref{e:lNhXLT_new}).  Therefore, we need LO input for the non-perturbative functions $D_1(z)$, $\tilde{g}(x)$, $g_T(x)$, and $g_1(x)$.\footnote{We note that several fits exist for the non-perturbative functions that enter our numerical study.  Given that we deal with LO formulas and are calculating an asymmetry, we believe the exact details of the parameterizations will not qualitatively affect our conclusions.} For $D_1(z)$ we use the DSS parameterization~\cite{deFlorian:2007aj} and for $g_1(x)$ we take the GRSV fit~\cite{Gluck:2000dy}. For $\tilde{g}(x)$, we look at two scenarios: i) using the approximate relation
\begin{equation}
\tilde{g}(x) = g_{1T}^{(1)}(x) \approx -f_{1T}^{\perp(1)}(x)\,, \label{e:Sivers}
\end{equation}
where the first equality was stated in Eq.~(\ref{e:g1T}), and we take the Sivers function from~\cite{Anselmino:2008sga};  and ii) using a WW-type approximation
\begin{equation}
\tilde{g}(x) \approx x\int_x^1\!\frac{dy} {y}g_1(y)\,, \label{e:gtilde}
\end{equation}
which was also used in Refs.~\cite{Kotzinian:2006dw,Avakian:2007mv,Kang:2011jw} and holds relatively well in certain models, like the quark model considered in Ref.~\cite{Pasquini:2008ax}. In both cases for $g_T(x)$ we use 
the WW approximation~\cite{Wandzura:1977qf},
\begin{equation}
g_T(x) \approx \int_x^1\!\frac{dy} {y}g_1(y)\,. \label{e:gT_WW}
\end{equation}
A few more comments are in order about the formula in Eq.~(\ref{e:Sivers}).  This relation is an approximation that can be roughly motivated by comparing $g_{1T}^{(1)}(x)$ in Ref.~\cite{Pasquini:2008ax} to $f_{1T}^{\perp(1)}(x)$ in Ref.~\cite{Pasquini:2010af}.  Also, even though the magnitudes of the ``worm-gear'' and Sivers functions are not the same, according to a large-$N_c$ analysis~\cite{Pobylitsa:2003ty}, one has $g^u_{1T} = -g^d_{1T}$ and $f^{\perp u}_{1T} = -f^{\perp d}_{1T}$. Therefore, the behavior of $f_{1T}^\perp$ qualitatively mimics that of $g_{1T}$, and we believe (\ref{e:Sivers}) is a worthwhile case to look at for this exploratory numerical study. 

\begin{figure}[t]
 \begin{center}
  \fig{0.52}{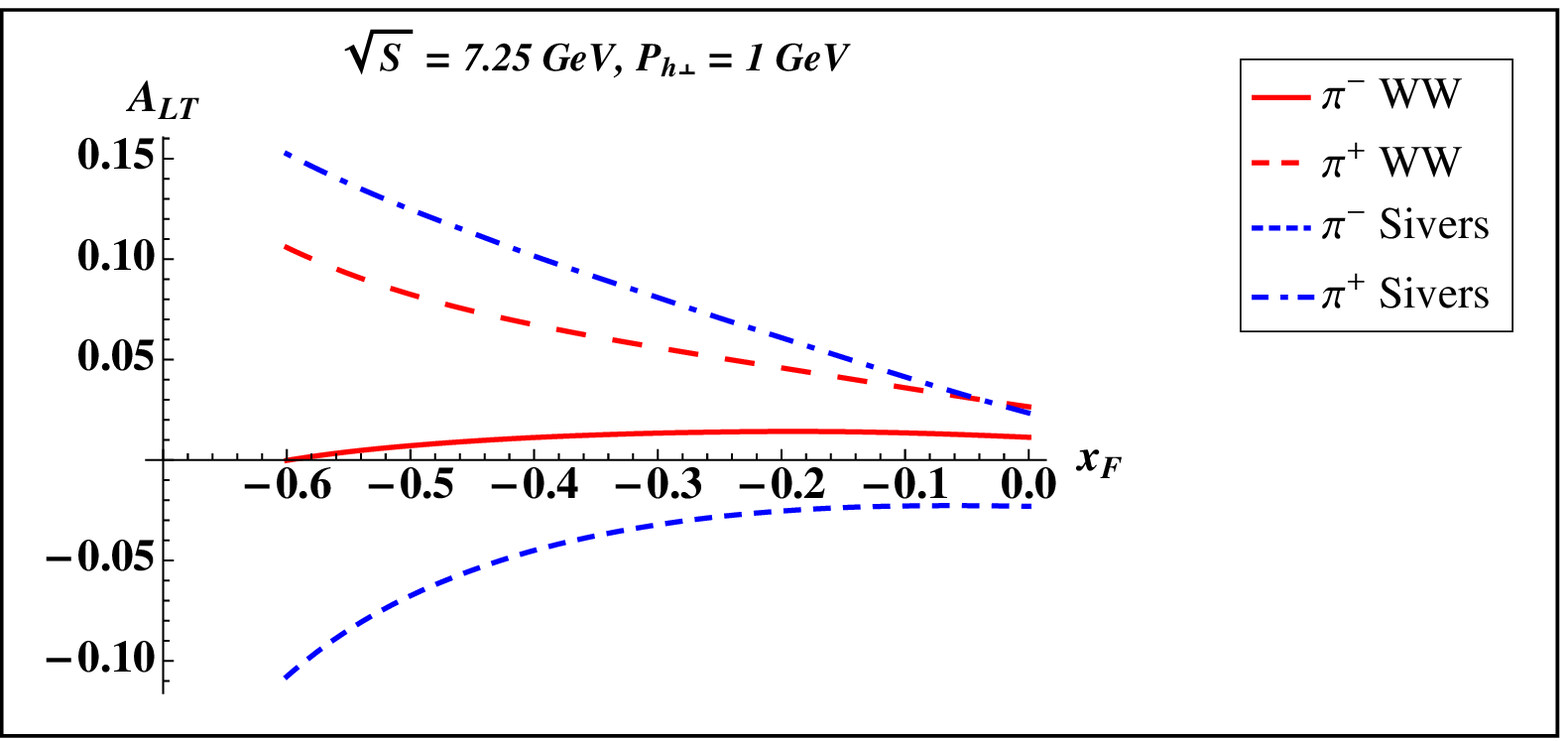}
  \fig{0.52}{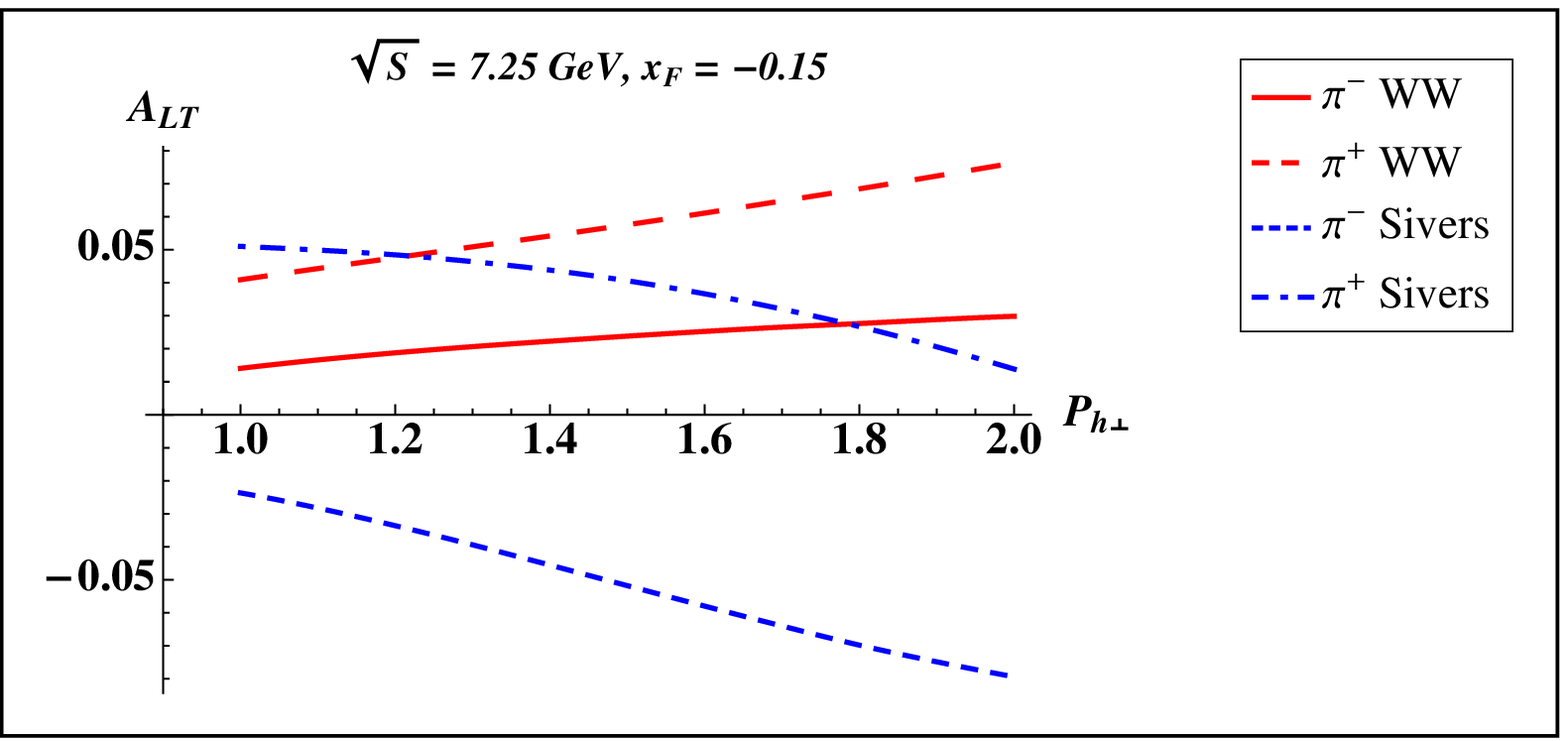}
 \end{center}
 \vspace{-0.65cm}
 \caption{$A_{LT}$ vs.~$x_F$ at fixed $P_{h\perp} = 1\,{\rm GeV}$ (left) and $A_{LT}$ vs.~$P_{h\perp}$ at fixed $x_F = -0.15$ (right) for HERMES $cm$ energy of $\sqrt{S} =7.25\,{\rm GeV}$. \label{f:HER}}
\end{figure}

\begin{figure}[t]
 \begin{center}
  \fig{0.52}{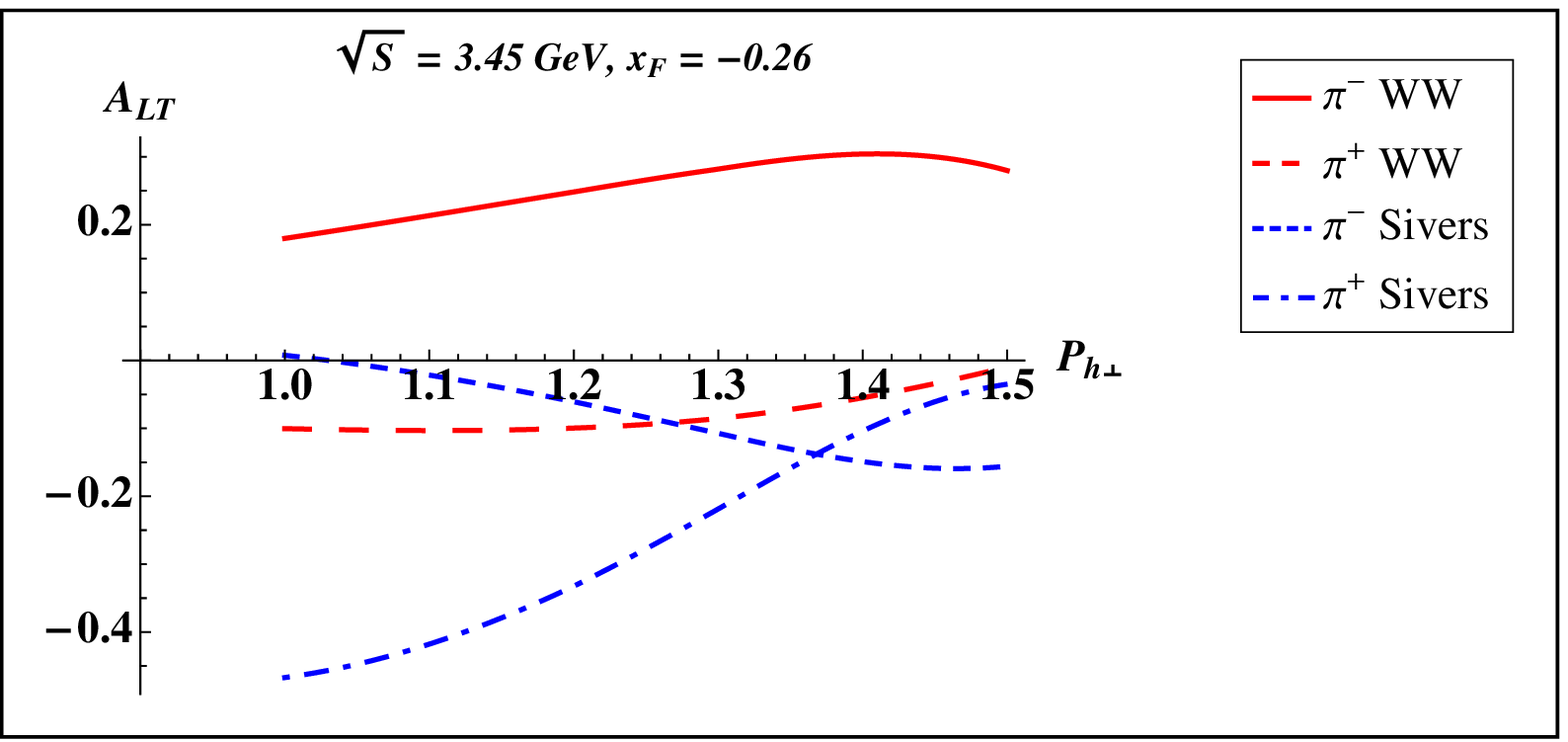}
  \fig{0.52}{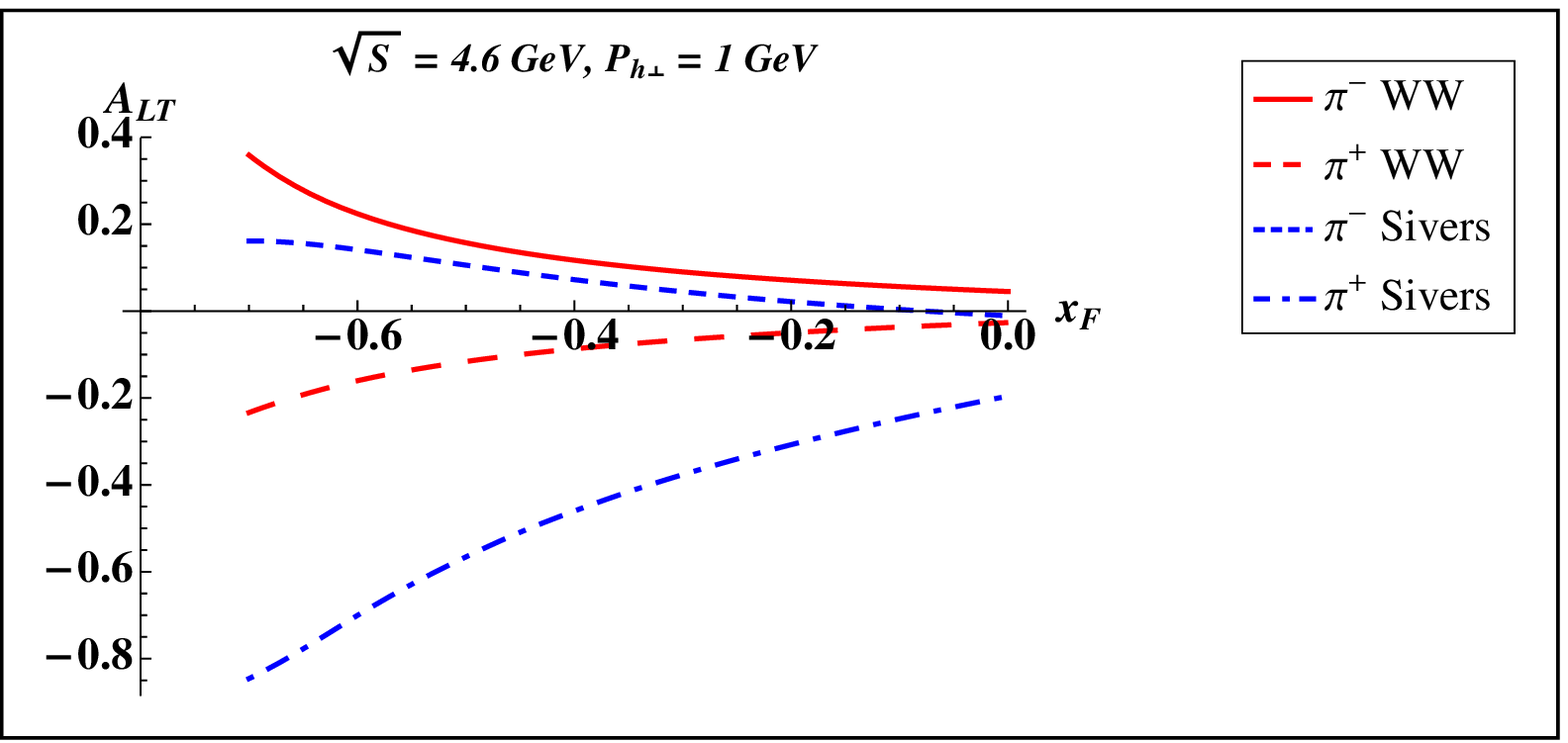}
 \end{center}
 \vspace{-0.65cm}
 \caption{$A_{LT}$ vs.~$P_{h\perp}$ at fixed $x_F = -0.26$ for JLab6 $cm$ energy of $\sqrt{S} =3.45\,{\rm GeV}$ (left) and $A_{LT}$ vs.~$x_F$ at fixed $P_{h\perp} = 1\,{\rm GeV}$ for JLab12 $cm$ energy of $\sqrt{S}=4.6\,{\rm GeV}$ (right). \label{f:JLab}}
\end{figure}

\begin{figure}[t]
 \begin{center}
  \fig{0.52}{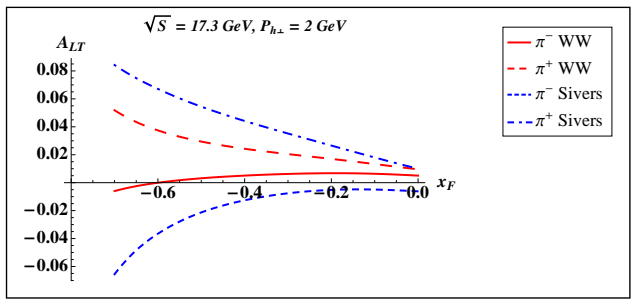}
 \end{center}
 \vspace{-0.65cm}
 \caption{$A_{LT}$ vs.~$x_F$ at fixed $P_{h\perp} = 2\,{\rm GeV}$ for COMPASS $cm$ energy of $\sqrt{S} =17.3\,{\rm GeV}$. \label{f:COM}}
\end{figure}

\begin{figure}[t]
 \begin{center}
  \fig{0.52}{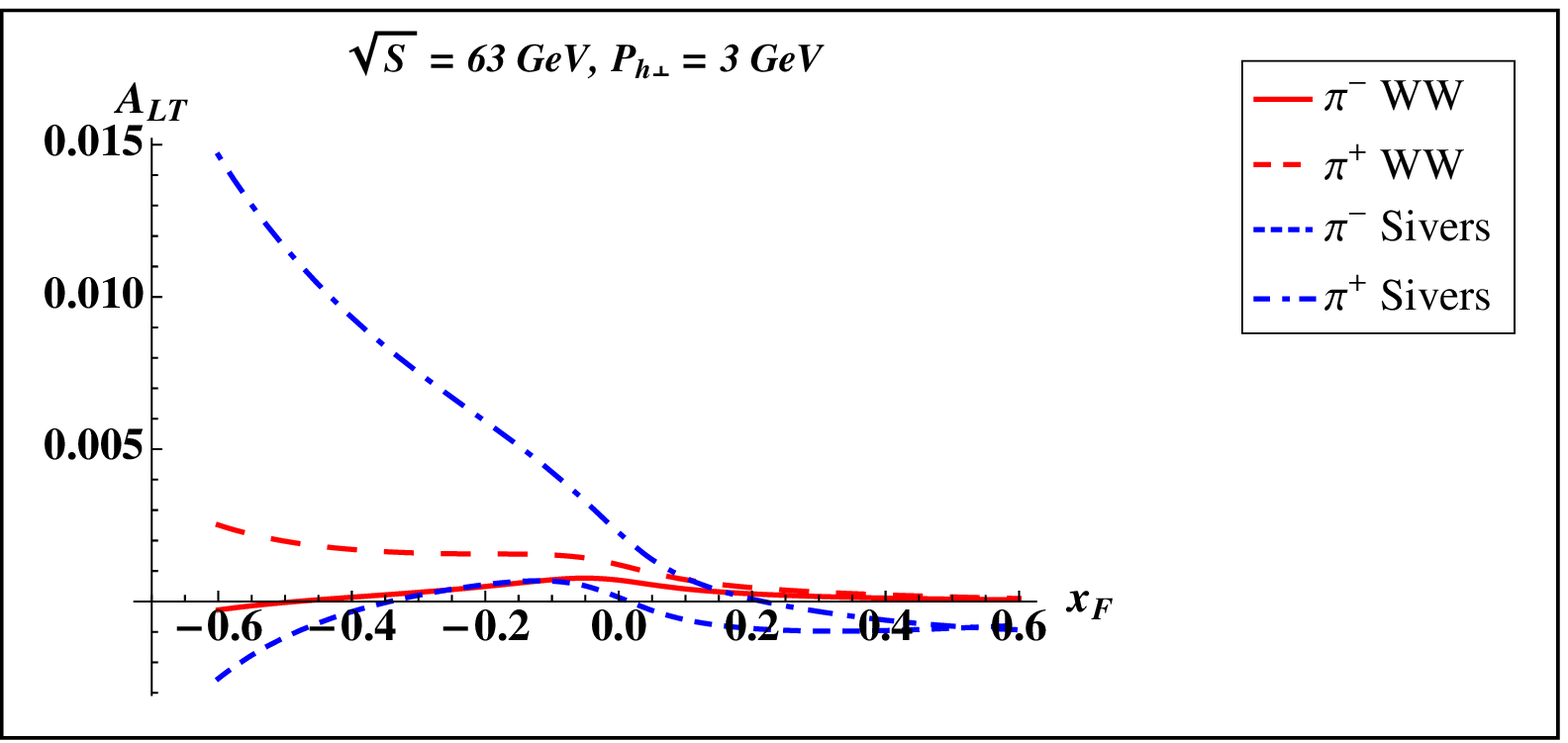}
  \fig{0.52}{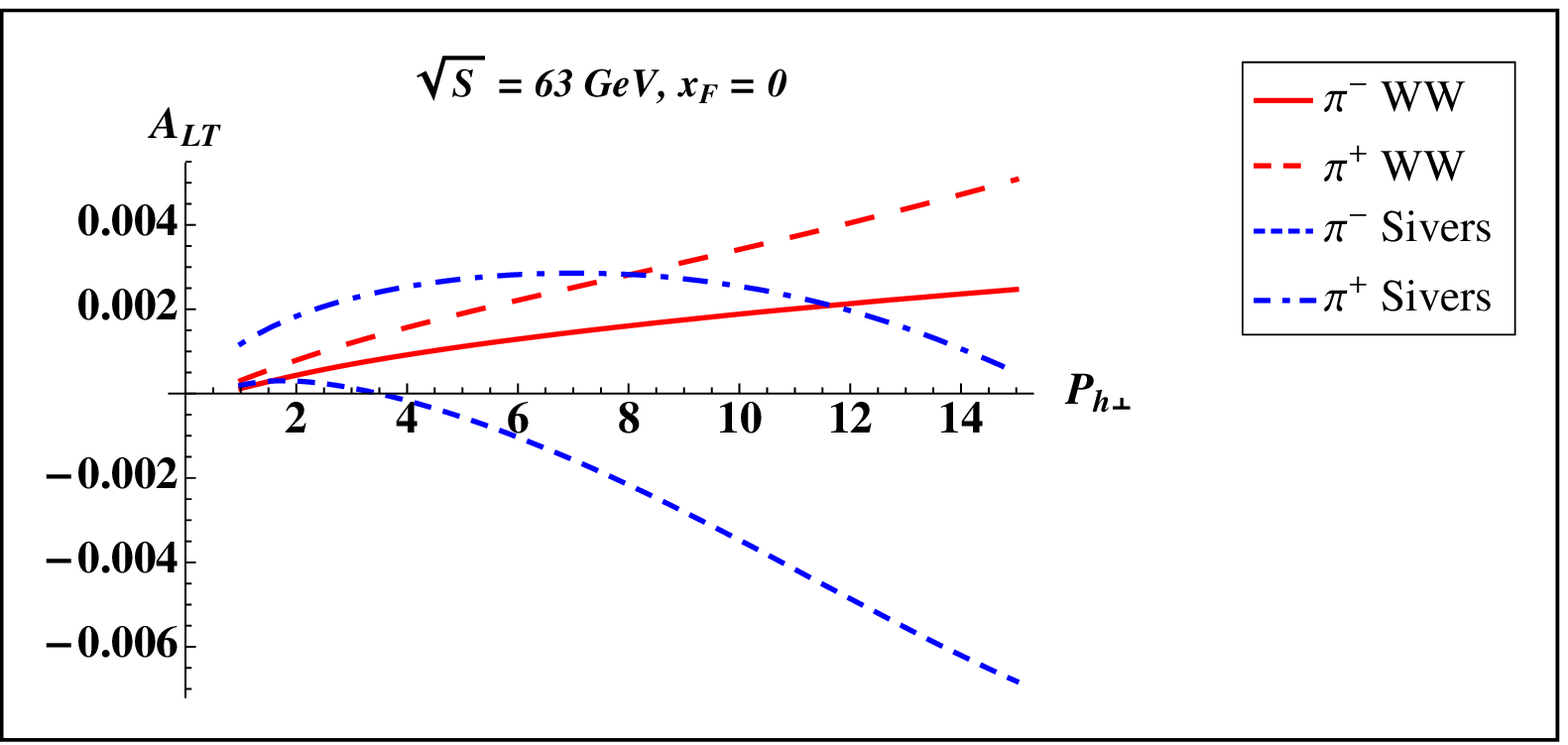}
 \end{center}
 \vspace{-0.65cm}
 \caption{$A_{LT}$ vs.~$x_F$ at fixed $P_{h\perp} = 3\,{\rm GeV}$ (left) and $A_{LT}$ vs.~$P_{h\perp}$ at fixed $x_F = 0$ (right) for EIC $cm$ energy of $\sqrt{S} =63\,{\rm GeV}$. \label{f:EIC}}
\end{figure}

\begin{figure}[t]
 \begin{center}
  \fig{0.515}{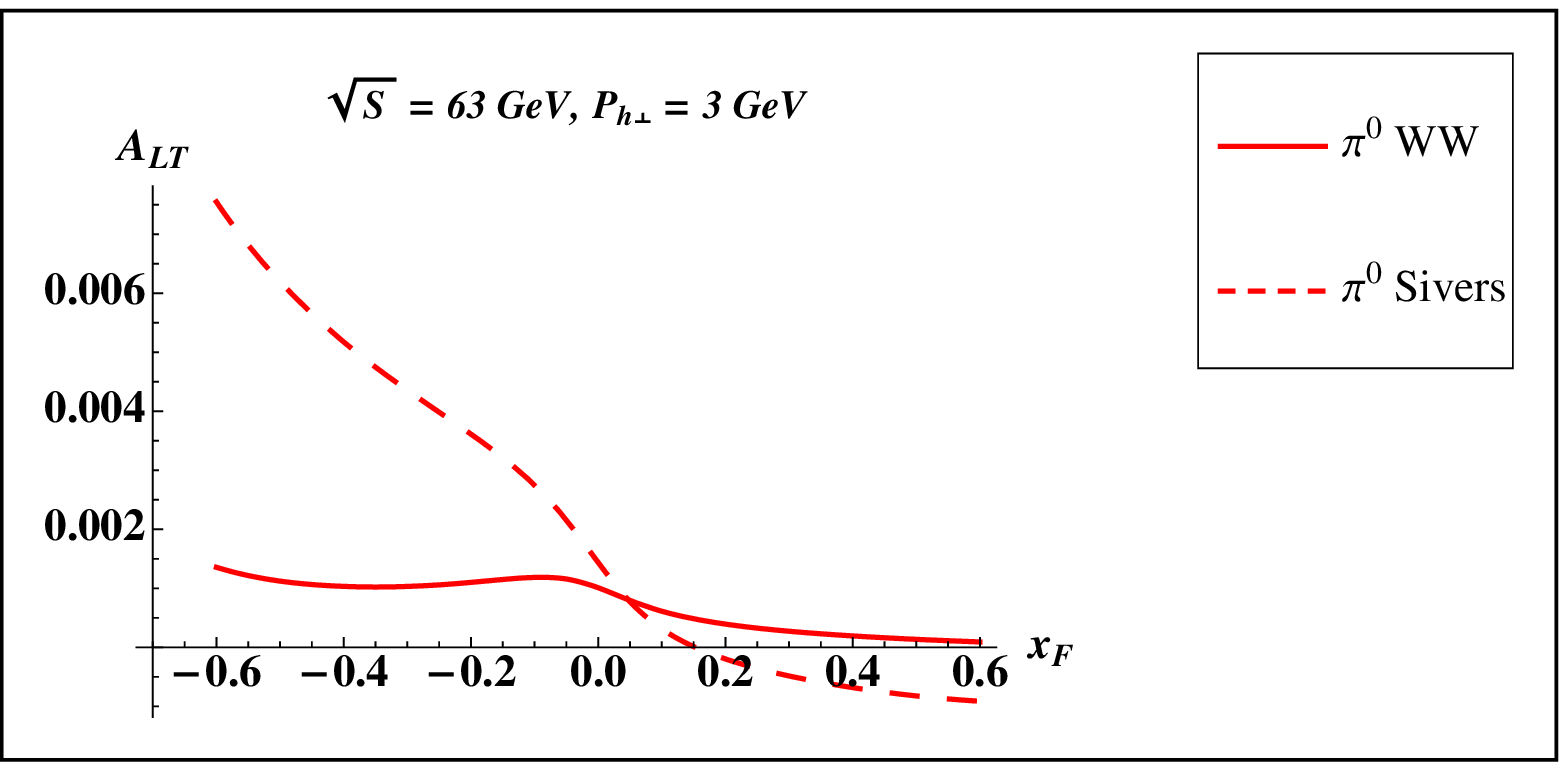}
  \fig{0.515}{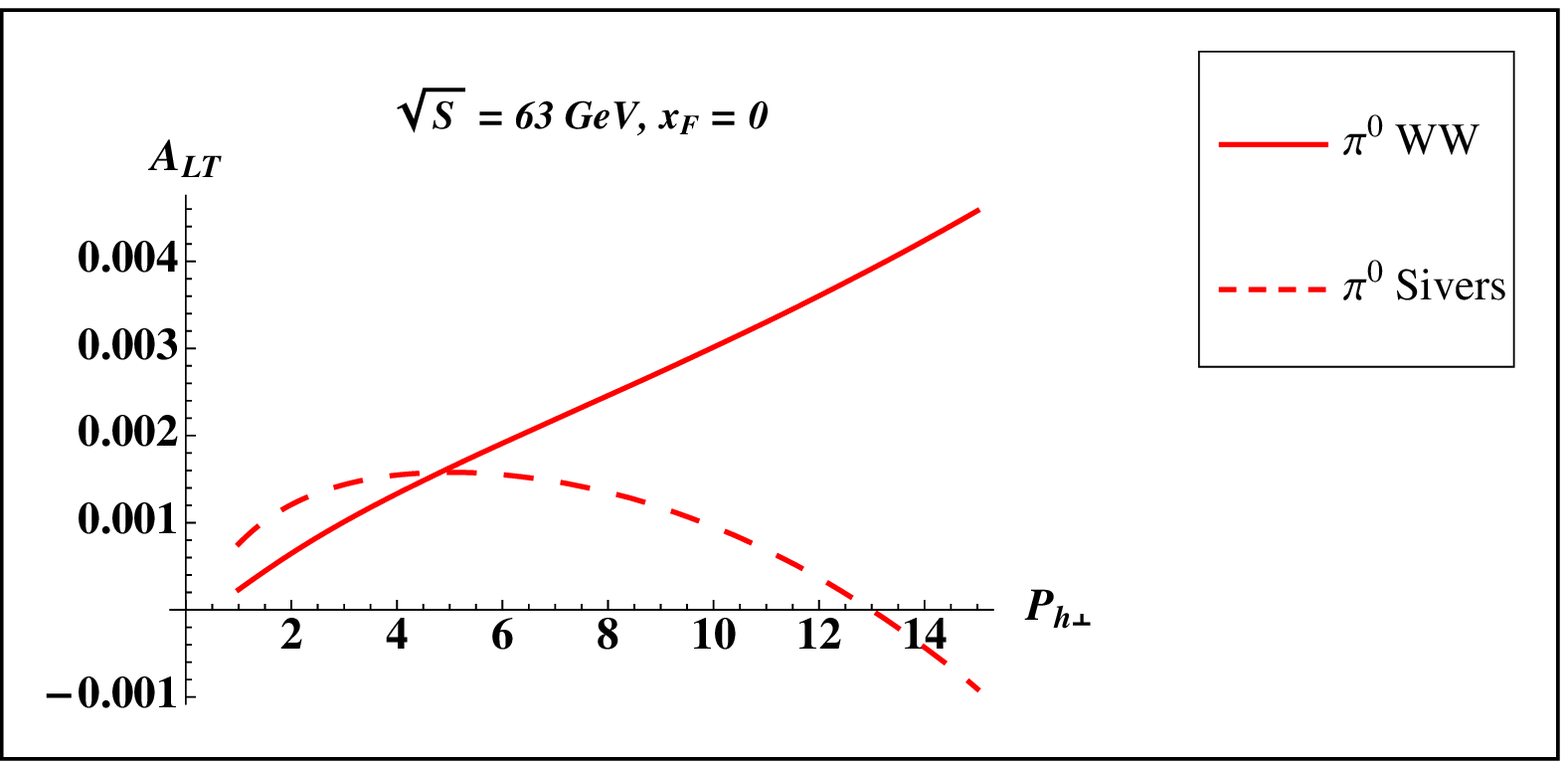}
 \end{center}
 \vspace{-0.65cm}
 \caption{Same as Fig.~\ref{f:EIC} but for neutral pions. \label{f:EIC0}}
\end{figure}

Here we give results for HERMES (Fig.~\ref{f:HER}), JLab6/JLab12 (Fig.~\ref{f:JLab}), COMPASS (Fig.~\ref{f:COM}), and EIC~(Figs.~\ref{f:EIC}, \ref{f:EIC0}) kinematics as functions of $x_F$ and/or $P_{h\perp}$.\footnote{We keep with the convention for $x_F$ used in transverse SSAs in proton-proton collisions, that is, $x_F > 0$ means hadrons detected in the direction of the transversely polarized proton.}  We mention that since the hadron is the only particle detected in the final state, its transverse momentum $P_{h\perp}$ sets the hard scale, and we require $P_{h\perp} \ge 1\,{\rm GeV}$.  All parton correlation functions are evaluated at the scale $P_{h\perp}$ with LO (DGLAP) evolution of the collinear functions.  We fix a lepton-nucleon $cm$ frame as described below Eq.~(\ref{e:LTreac}), with $A_{LT}$ defined as in Eq.~(\ref{e:ALT}).  The ``Sivers'' curves are found using Eq.~(\ref{e:Sivers}), whereas the ``WW'' curves come from using Eq.~(\ref{e:gtilde}).  One immediately notices that in all the plots the Sivers and WW curves can be quite different and sometimes not even have the same sign.  This is due to the fact that $\tilde{g}^{d/p}(x)$ is much smaller in the WW scenario than in the Sivers scenario, while $\tilde{g}^{u/p}(x)$ is similar for the two cases (see Fig.~\ref{f:functions}).\footnote{We note that the only difference between the WW and Sivers scenarios comes from $\tilde{g}(x)$ because $g_T(x)$ and $g_1(x)$ (also shown in Fig.~\ref{f:functions}) are the same for both cases.}  Thus, for a proton target (see Figs.~\ref{f:HER}, \ref{f:COM}--\ref{f:EIC0}) one finds $A_{LT}$ for $\pi^+$ production is similar (at least of the same sign) for the WW and Sivers scenarios, while for $\pi^-$ production the two cases are not alike and do not even have the same sign.  This occurs because in the WW scenario, the smallness of $\tilde{g}^{d/p}(x)$, along with the weight of 1/9 from the down quark charge (squared), allows the up quark contribution (even though it contains a disfavored FF) to ``overcome'' the down quark term and make $A_{LT}$ positive.  However, in the Sivers case, $\tilde{g}^{d/p}(x)$ is of similar magnitude (but opposite in sign) to $\tilde{g}^{u/p}(x)$, so one obtains a negative $A_{LT}$ for $\pi^-$ production (compared to a positive asymmetry for $\pi^+$).  

For the neutron, one can make a na\"{i}ve argument on what to expect for $A_{LT}$ based on isospin relations, quark charge (squared) weights, and the relative sizes of $\tilde{g}(x)$ for down and up quarks and $D_1(z)$ for favored and disfavored fragmentation.  For the Sivers scenario, one finds both $\pi^+$ and $\pi^-$ are driven by $\tilde{g}^{u/n}(x)$.  Thus, both have the same sign, but with $\pi^-$ smaller in magnitude due to a factor of the disfavored FF.  For the WW case, $A_{LT}$ for $\pi^+$ is also driven by $\tilde{g}^{u/n}$, while for $\pi^-$ the asymmetry is mainly due to $\tilde{g}^{d/n}(x)$.  As a result, the former has the same sign as the Sivers case, while the latter has the opposite sign (see the left panel of Fig.~\ref{f:JLab}).  However, this argument can be spoiled by the kinematical dependence of the cross section and the fact that for the neutron target one has a non-negligible contribution from $g_1(x)$ (see the right panel of Fig.~\ref{f:gtilgTg1}), as seen in the right panel of Fig.~\ref{f:JLab}.  Therefore, one has to take into account the details on the experiment before making any definite statements on what to expect from the WW and Sivers scenarios.  In general, one also has to keep in mind that for the Sivers input there are errors~\cite{Anselmino:2008sga} and the ``range'' spanned by different Sivers curves could give an improvement in the agreement one finds with the WW scenario.

Overall, measurements of $A_{LT}$ in this reaction might help distinguish between the two scenarios considered here, where even the sign of the asymmetry could give a first indication on the form of $\tilde{g}(x)$.  It was already found in quark model calculations that the WW-type approximation in Eq.~(\ref{e:gtilde}) should be a decent approximation to the full function~\cite{Pasquini:2008ax}.  Experiments should help to confirm/refute this.  Should the magnitude of the data be significantly different from our numerical predictions, one might conjecture that quark-gluon-quark correlations in the nucleon and/or twist-3 fragmentation effects (i.e., Eq.~(\ref{e:lNFrag})) are important. On the other hand, if one finds results comparative to ours, then such effects could be excluded.  This could then allow for a ``clean'' extraction of $\tilde{g}(x)$, due to the $g_1(x) + g_T(x)$ piece being extremely small for a proton target and the $g_T(x)$ contribution (for the most part) being negligible for a neutron target (see Fig.~\ref{f:gtilgTg1}).  Moreover, one has the best chance make such an extraction using data from HERMES, JLab, and COMPASS since once one moves towards a higher $cm$ energy (i.e., an EIC), the asymmetry becomes very small (see Figs.~\ref{f:EIC}, \ref{f:EIC0}), which was also seen in $A_{LT}$ in jet production~\cite{Kang:2011jw}.  However, as was emphasized in Ref.~\cite{Gamberg:2014eia}, at the relatively low $P_{h\perp}$ of HERMES, JLab, and COMPASS, due to quasi-real photoproduction, one would most likely need a NLO calculation to make any rigorous quantitative conclusions. 

\begin{figure}[t]
 \begin{center}
  \fig{0.52}{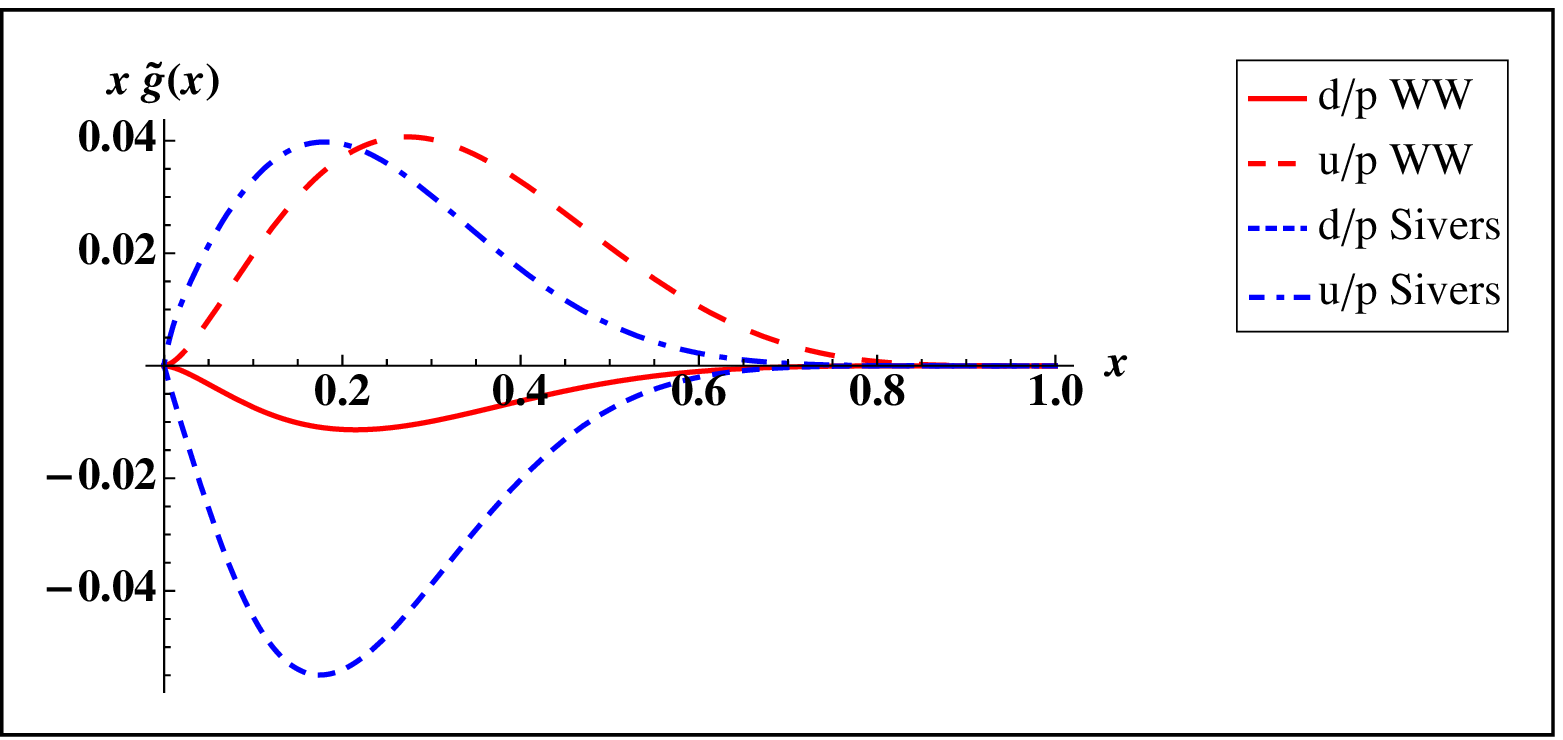}
  \fig{0.52}{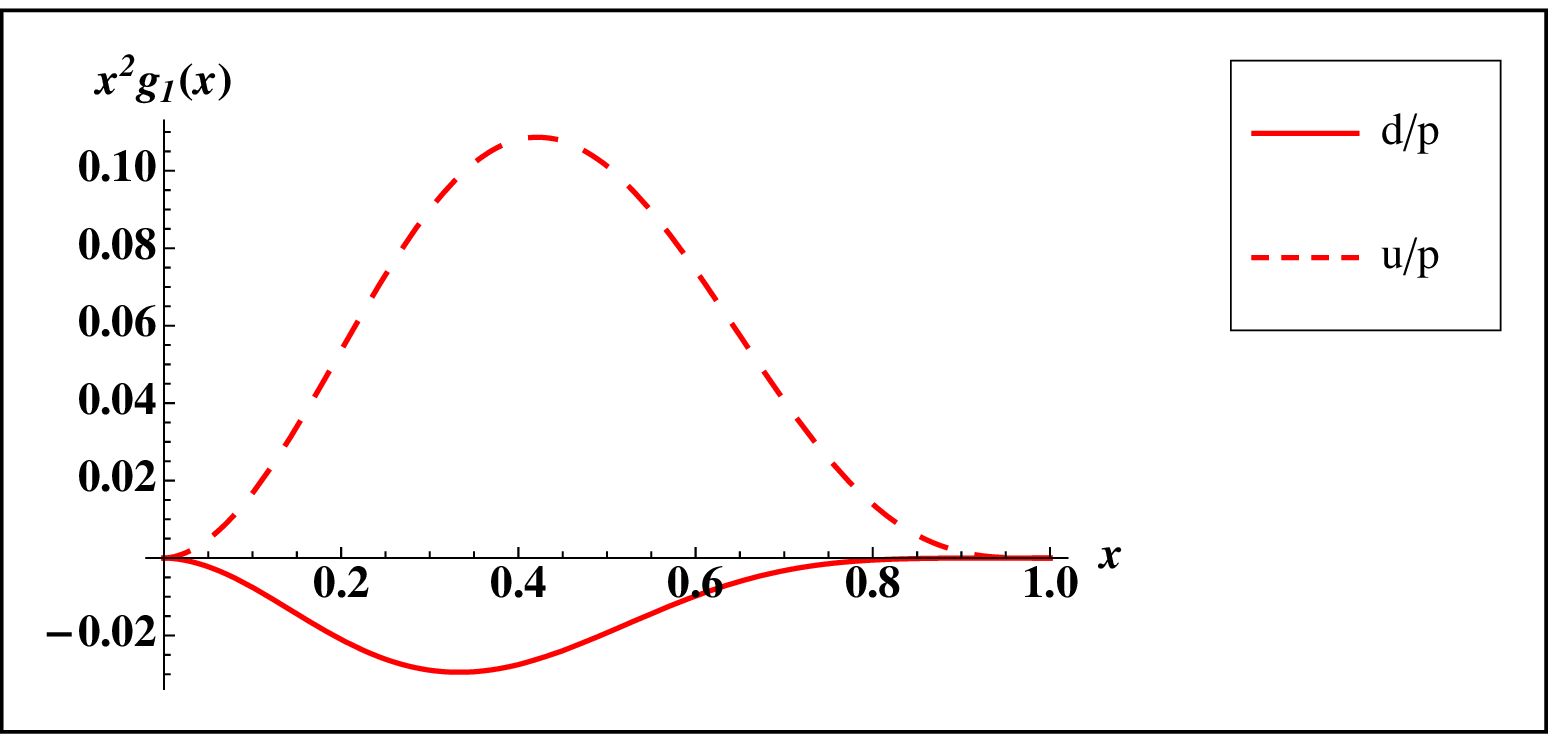}
 \end{center}
 \vspace{-0.65cm}
 \caption{Plots of $x\,\tilde{g}(x)$ vs.~$x$ (left) and $x^2\,g_1(x)$ vs.~$x$ (right) at a scale $\mu = 2\,{\rm GeV}$ for down and up quarks in a proton.  For $\tilde{g}(x)$ we give both the WW and Sivers scenarios, while $g_1(x)$ is the same in both cases. We multiply $g_1(x)$ by $x^2$ instead of $x$ since this function appears in the cross section (\ref{e:lNhXLT_new}) with a factor of $x$ compared to $\tilde{g}(x)$.  Note that $x^2\,g_T(x)$ vs.~$x$ is identical to the $x\,\tilde{g}(x)$ plot in the WW scenario.\label{f:functions}}
\end{figure}

\begin{figure}[t]
 \begin{center}
  \fig{0.55}{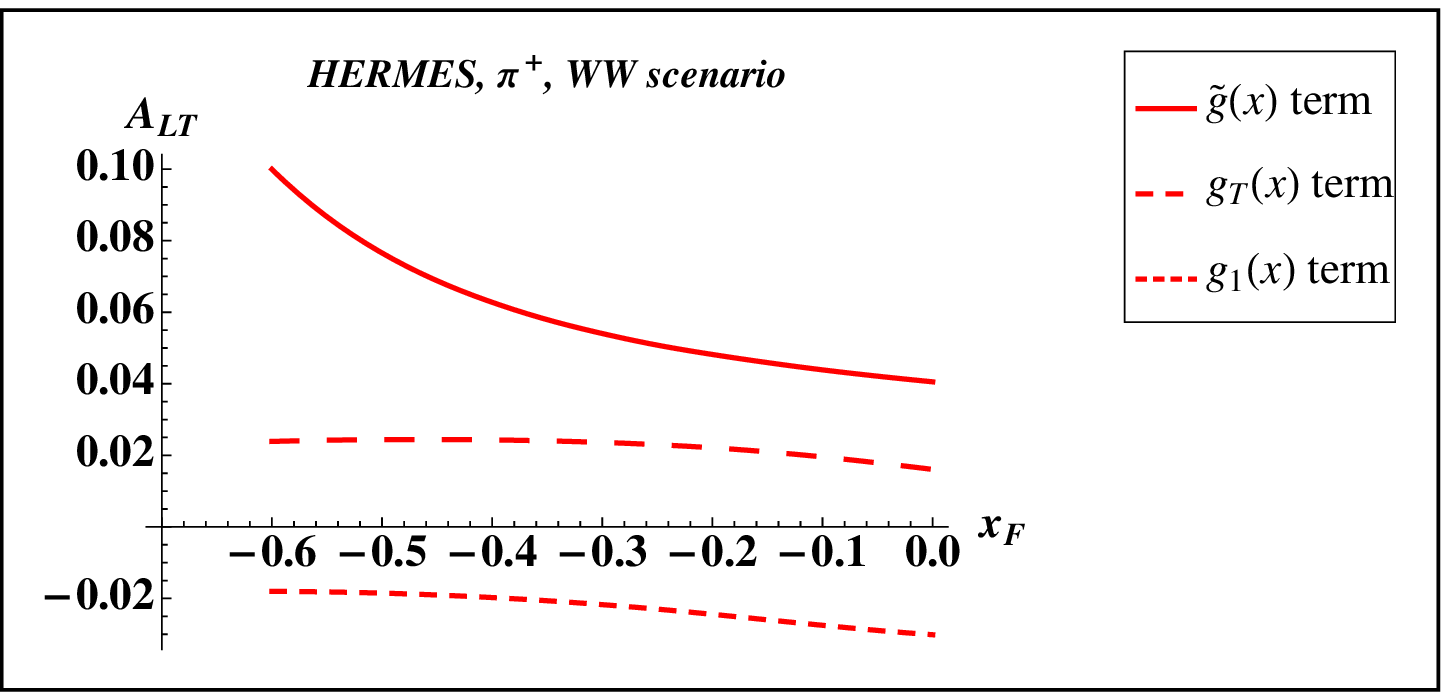}
  \fig{0.56}{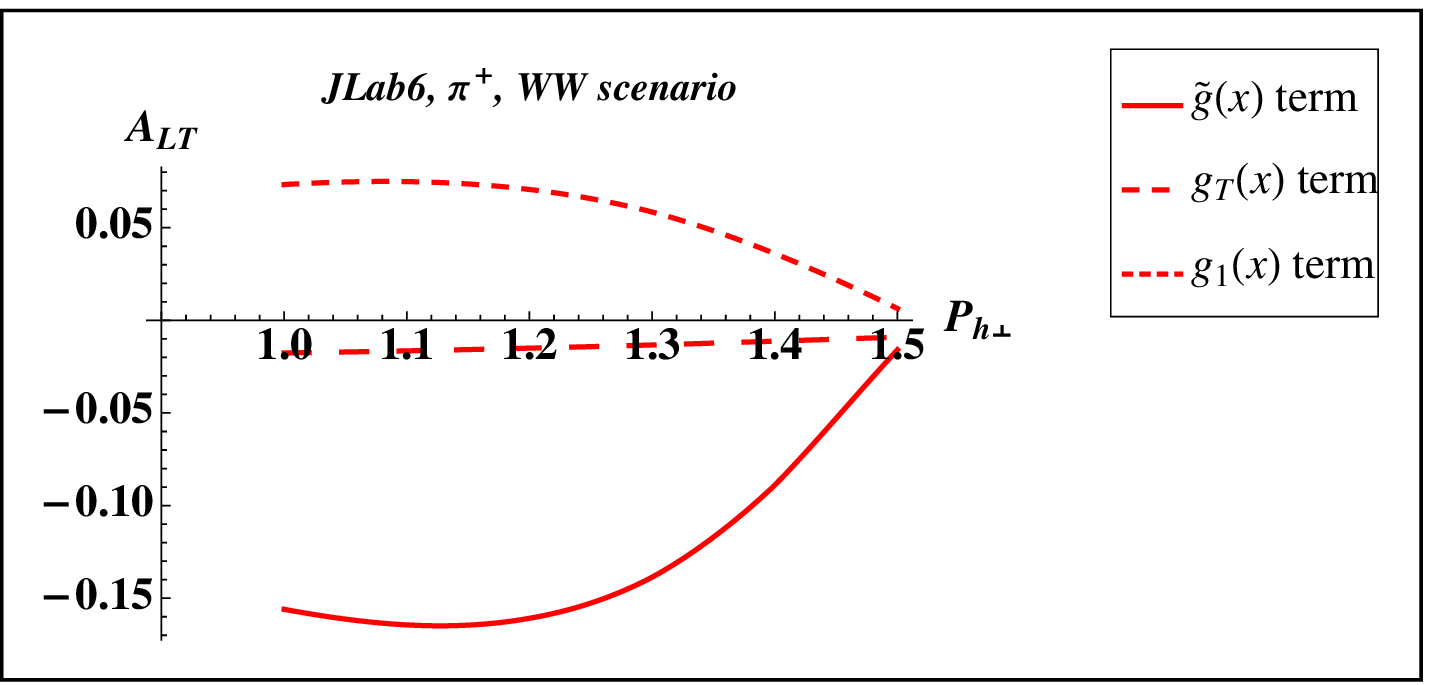}
 \end{center}
 \vspace{-0.65cm}
 \caption{Individual contributions to $A_{LT}$ from the $\tilde{g}(x)$, $g_T(x)$, and $g_1(x)$ terms in Eq.~(\ref{e:lNhXLT_new}) for HERMES (left) and JLab6 (right) kinematics.  These plots correspond to the left panels of Fig.~\ref{f:HER} and Fig.~\ref{f:JLab}, respectively, for $\pi^+$ production in the WW scenario.  Similar conclusions hold for $\pi^-$ and the Sivers case, with the exception that at JLab6 kinematics for $\pi^+$ production in the Sivers scenario, the $\tilde{g}(x)$ term dominates at low $P_{h\perp}$, and for $\pi^-$ production in that same case, the $g_T(x)$ piece at low $P_{h\perp}$ is somewhat comparable to the other terms. \label{f:gtilgTg1}}
\end{figure}

%
%
\section{Summary} \label{s:sum}
In this paper we have analyzed the longitudinal-transverse double-spin asymmetry in single inclusive leptoproduction of hadrons within the framework of collinear twist-3 factorization.  Both HERMES and JLab Hall A are expected to have results on this observable, and this effect can also be measured at COMPASS and a future EIC.  We provided an analytical result for the double-spin dependent cross section, including both the distribution and fragmentation terms, as well as a phenomenological study of the former using known non-perturbative inputs.  For $A_{LT}$ we looked at two scenarios for the twist-3 function $\tilde{g}(x)$: a ``Sivers'' input and a ``Wandzura-Wilczek'' input.  We found that these two cases can give quite different results due to the different behavior of $\tilde{g}(x)$.  Thus, even a qualitative comparison of our predictions with experiment could help distinguish between the Sivers and WW scenarios.  Moreover, if the magnitude of the data is in line with our results, one could have direct access to the ``worm-gear''-type function $\tilde{g}(x)$, which plays a role in certain azimuthal asymmetries in SIDIS and has gained interest over the years.  If the magnitude is not in agreement, this observable could give insight into the importance of quark-gluon-quark correlations in the nucleon and/or twist-3 fragmentation effects in unpolarized hadrons. However, one always has to keep in mind the potential large impact of NLO terms. In general, we found the best chance to measure a nonzero asymmetry is at HERMES, JLab, and COMPASS, as the high $cm$ energy of an EIC leads to a very small effect.  We expect this conclusion to be rather robust upon including higher order corrections.

\section*{Acknowledgments}

A.M. would like to thank D.~Flay, Z.-E.~Meziani, and M.~Posik for discussions about the preliminary data on $A_{LT}$ from JLab, and D.P.~appreciates a useful conversation with X.~Jiang on this data as well.  This work has been supported by the National Science
Foundation under Contract No.~PHY-1205942 (K.K. and A.M.), and the RIKEN BNL
Research Center (D.P.).

%
%
%
%
\section*{Appendix A: Frame-independence of Eq.~(\ref{e:lNhXLT_new})}

In this appendix we show how one can obtain the same result for the
double-spin dependent cross section in the lepton-nucleon and nucleon-hadron $cm$ frames. 
To this end, it is convenient to write the cross section in a
manifestly covariant way. A straightforward twist-3 calculation in a
general frame gives 
\begin{align}
&\hspace{-0.3cm}\frac{P_h^0\,d\sigma_{LT}^{Dist}(\lambda_\ell,\vec{S}_\perp)} {d^3\vec{P}_h} =
 -\frac{8\alpha_{em}^2}
 {S}\,M\,\lambda_\ell\sum_q
 e_q^2\int_{z_{min}}^1\!\frac{dz} {z^3}\,\frac{1} {S+T/z}\,\frac{1}
 {x\hat{u}}\,D_1^{h/q}(z)\nonumber\\[0.05cm]
&\hspace{2cm}\times\,\Bigg\{\!
 \,g_1^q(x)\!\left[\frac{z\hat{u}(\hat{s}-\hat{u})} {4 
 \hat{t}}\right]
 (n\cdot S) 
+ \left(\tilde{g}^q(x)-x\frac{d\tilde{g}^q(x)}
 {dx}\right)\!
 \left[ \frac{\hat{u}-\hat{s}} {2\hat{t}^{\hspace{0.025cm}2}} \right]
 \left( \hat{s} (P_h\cdot\Sp) + z\hat{t} (l\cdot\Sp) \right)\nonumber\\[0.05cm]
&\hspace{3cm} + x\,g_T^q(x)\left[\frac{\hat{u}}
 {2\hat{t}}\right] \left( - P_h\cdot\Sp + 2 z (l\cdot \Sp) \right)
 + \int
 \!dx_1\,G_{DT}^q(x,x_1)\left[ \frac{\hat{u}(\hat{u}-\hat{s}) \,
 P_h\cdot\Sp}
 {\xi\hat{t}^{\hspace{0.025cm}2}}\right]  \!\Bigg\}\,,  \label{e:gen}
\end{align}
where $\Sp^\mu = S^\mu - (n\cdot S) P^\mu$ with $n^\mu$ being a lightlike
vector satisfying $P\cdot n=1$. Here we include the contribution from
the twist-2 quark helicity distribution $g_1(x)$ since $n\cdot S$ survives the asymmetry (\ref{e:ALT}) in any frame where the momenta of
the initial lepton and nucleon are not collinear.  Using the LIR (\ref{e:LIR}) in our formula (\ref{e:gen}), we can eliminate the 3-parton correlator to obtain
\begin{align}
&\hspace{-0.3cm}\frac{P_h^0\,d\sigma_{LT}^{Dist}(\lambda_\ell,\vec{S}_\perp)} {d^3\vec{P}_h} =
 -\frac{8\alpha_{em}^2}
 {S}\,M\,\lambda_l\sum_q
 e_q^2\int_{z_{min}}^1\!\frac{dz} {z^3}\,\frac{1} {S+T/z}\,\frac{1}
 {x\hat{u}}\,D_1^{h/q}(z)\nonumber\\[0.05cm]
&\hspace{2cm}\times\,\Bigg\{\!
 \,x \, g_1^q(x)\! \left[ \frac{\hat{u}(\hat{s}-\hat{u})} {4 
 \hat{t}^2} \right] \left( \frac{z\hat{t}}{x}(n\cdot S) - 2 P_h\cdot\Sp
 \right)\nonumber\\[0.05cm]
&\hspace{3cm}+ \left(\tilde{g}^q(x)-x\frac{d\tilde{g}^q(x)}
 {dx}\right)\!
 \left[ \frac{\hat{u}-\hat{s}} {2\hat{t}^{\hspace{0.025cm}2}} \right]
 \left( \hat{s} (P_h\cdot\Sp) + z\hat{t} (l\cdot\Sp) \right)\nonumber\\[0.05cm]
&\hspace{3cm} + x\,g_T^q(x)\left[\frac{\hat{u}}
 {\hat{t}}\right] \left( \frac{\hat{s}} {\hat{t}} (P_h\cdot\Sp) + z (l\cdot \Sp) \right) \!\Bigg\}.
 \label{e:gen2}
\end{align}
We now demonstrate how the cross section (\ref{e:gen2}) leads to the same result in the two
frames.
In the lepton-nucleon $cm$ frame, one has $n^\mu=l^\mu/(P\cdot l)$, whereas in the nucleon-hadron $cm$ frame, one has $n^\mu = P_h^\mu/(P\cdot P_h)$. 
The scalar products in Eq.~(\ref{e:gen2}) can be written as
 \begin{align}
 l\cdot S_\perp\,\vb_{n=\frac{l}{P\cdot l}}  &= 0\,,\label{e:scal1}\\
 n\cdot S \,\vb_{n=\frac{P_h}{P\cdot P_h}} &= 
  \left( -\frac{2x}{z\hat{t}}\,P_h\cdot\Sp + \frac{l\cdot S}{P\cdot l}
    \right)\!
  \vb_{n=\frac{l}{P\cdot l}}\,, \label{e:scal2}\\
 P_h\cdot\Sp\,\vb_{n=\frac{P_h}{P\cdot P_h}} &= 0\,, \label{e:scal3}\\
 l\cdot\Sp\,\vb_{n=\frac{P_h}{P\cdot P_h}} &= \frac{\hat{s}}{z\hat{t}}
 \,P_h\cdot\Sp\,\vb_{n=\frac{l}{P\cdot l}}\,. \label{e:scal4}
 \end{align}
Note that the scalar product $l\cdot S$ in the lepton-nucleon $cm$ frame does
not contain $\Sp$.  Therefore, any term proportional to $l\cdot S$ will not survive the asymmetry (\ref{e:ALT}), so one can ignore such pieces.  From these relations, the frame-independence of the cross section becomes manifest.  That is, the expression in Eq.~(\ref{e:gen2}) can be evaluated in the lepton-nucleon $cm$ frame by means of Eq.~(\ref{e:scal1}) or in the nucleon-hadron $cm$ frame by means of Eqs.~(\ref{e:scal2})--(\ref{e:scal4}). In either case, one easily sees the same hard scattering coefficients show up for each function, which leads to the result in Eq.~(\ref{e:lNhXLT_new}).

\end{document}